\documentclass[aps,prd,twocolumn,10pt]{revtex4-1}

\usepackage{ifthen}
\usepackage{hyperref}
\usepackage{graphicx}
\usepackage{amsmath}
\usepackage{color}
\usepackage[T1]{fontenc}

\begin{document}

\title{Is there another coincidence problem at the reionization epoch?}

\author{Lucas~Lombriser}
\affiliation{Institute for Astronomy, University of Edinburgh, Royal Observatory, Blackford Hill, Edinburgh, EH9~3HJ, U.K.}
\author{Vanessa Smer-Barreto}
\affiliation{Institute for Astronomy, University of Edinburgh, Royal Observatory, Blackford Hill, Edinburgh, EH9~3HJ, U.K.}

\date{\today}


\begin{abstract}

The cosmological coincidences between the matter and radiation energy densities at recombination as well as between the densities of matter and the cosmological constant at present time are well known.
We point out that moreover the third intersection between the energy densities of radiation and the cosmological constant coincides with the reionization epoch.
To quantify the statistical relevance of this concurrence, we compute the Bayes factor between the concordance cosmology with free Thomson scattering optical depth and a model for which this parameter is inferred from imposing a match between the time of density equality and the epoch of reionization.
This is to characterize the potential explanatory gain if one were to find a parameter-free physical connection.
We find a \emph{very strong} preference for such a concurrence on the Jeffreys scale from current cosmological observations.
We furthermore discuss the effect of choice of priors, changes in reionization history, and free sum of neutrino masses.
We also estimate the impact of adding intermediate polarization data from the \emph{Planck} High Frequency Instrument and prospects for future 21~cm surveys. In the first case, preference for the correlation remains \emph{substantial}, whereas future data may give results more \emph{decisive} in \emph{pro} or \emph{substantial} in \emph{contra}.
Finally, we provide a discussion on different interpretations of these findings.
In particular, we show how a connection between the star-formation history and the cosmological background dynamics can give rise to this concurrence.

\end{abstract}


\maketitle


\section{Introduction} \label{sec:intro}

The observed late-time accelerated expansion of our Universe~\cite{Riess:1998cb,Perlmutter:1998np} remains a difficult and enduring puzzle to modern physics.
It is generally attributed to a cosmological constant $\Lambda$, thought to arise from vacuum fluctuations.
Quantum theoretical calculations are, however, off by $\gtrsim50$ orders of magnitude from the observed value~\cite{Weinberg:1988cp,Martin:2012bt} (cf.~\cite{Wang:2017oiy}).
Cosmic acceleration could instead be driven by a dark energy field that permeates the Universe or be due to a breakdown of general relativity on large scales~\cite{Clifton:2011jh,Koyama:2015vza,Joyce:2016vqv}, perhaps with an unrelated mechanism suppressing the vacuum contributions.
Dark energy must closely mimic a cosmological constant however (e.g.,~\cite{Ade:2015xua}),
and the combination of large-scale structure and gravitational wave observations poses a challenge to the concept of cosmic self acceleration from a genuine modification of gravity~\cite{Lombriser:2014ira,Jimenez:2015bwa,Lombriser:2015sxa,Brax:2015dma,Lombriser:2016yzn}.

An enigmatic aspect of the cosmological constant problem is the coincidence that the energy densities of the cosmological constant and matter happen to be comparable in size at the present era.
This \emph{Why Now?} problem may be used as a guide in the search for possible solutions to the conundrum around cosmic acceleration~\cite{Martin:2012bt}.
Among many other approaches, this involves anthropic and multiverse ideas~\cite{Carter:1983,Weinberg:1987dv,Peacock:1999ye} (e.g., $\Lambda$ dependence of star-formation rate~\cite{Lineweaver:2007sla,Bousso:2008bu,Loeb:2016vps}) or nonlocal concepts~\cite{Kaloper:2013zca}.
Other cosmological coincidences revolve around the comparable sizes of the baryonic and dark matter energy densities or the equality of the energy densities of radiation and matter at the recombination epoch~\cite{Peacock:1999ye}. Recombination occurred as the first of two universal phase transitions of the baryonic gas at redshift $z\sim1100$ when the Universe had cooled sufficiently to allow electrons and protons to form neutral hydrogen~\cite{Peebles:1968,Zeldovich:1969,Seager:1999km}. This enabled photon-matter decoupling and produced the observed last-scattering surface of the cosmic microwave background (CMB).
For a universe with different parameters this would not obviously need to concur with the matter-radiation equality.

A seemingly unrelated problem is the physics underlying the second universal phase transition of the gas, the epoch of reionization~\cite{Barkana:2000fd,Zaroubi:2012in,Mesinger:2016}, commencing with the end of the \emph{Dark Ages}.
While the processes involved are not yet understood in full detail, this ionization of the hydrogen and helium gas is attributed to the radiation from early galaxies ($z\sim12-6$) and quasars ($z\sim6-2$).
These objects ionized the intergalactic medium surrounding them and after becoming sufficiently abundant eventually ionized the entire Universe.
Whether additional sources are needed to explain the observed reionization is, however, still being investigated~\cite{Barkana:2000fd,Zaroubi:2012in,Mesinger:2016}.
The optical depth due to Thomson scattering of CMB photons in the ionized Universe can be observed in the large-angle CMB polarization anisotropies~\cite{Ade:2015xua,Aghanim:2016yuo,Adam:2016hgk}.
The lower redshift limit for reionization is inferred from the Gunn-Peterson~\cite{Gunn:1965} effect in the absorption spectra of quasar and gamma ray burst radiation interacting with the intergalactic medium~\cite{Becker:2001ee,Fan:2005eq,Venemans:2013npa,Becker:2015lua}.
Similarly, Lyman-$\alpha$ emissions can be used to give an upper bound on the reionization redshifts~\cite{Tilvi:2014oia,Schenker:2014tda,Fan:2005eq}.
While limited information is available today, ongoing and future 21~cm surveys will unveil a lot more details of the processes governing reionization~\cite{Zaroubi:2012in,Liu:2015txa}.

With the cosmological coincidences of the energy densities of matter and radiation intersecting at recombination and of the present equality in the densities of matter and the cosmological constant, it seems natural to inspect the third intersection where the cosmological constant and radiation densities are equal.
In this paper, we point out for the first time that this third equality coincides with the epoch of reionization.
We conduct a statistical analysis of this concurrence in light of current and future cosmological data and provide a discussion on whether this observed correlation should be considered a \emph{new cosmological coincidence problem}.
We also view the concurrence in terms of the star-formation history to the epoch of reionization.

The paper is organized as follows.
In Sec.~\ref{sec:coincidence}, we briefly review the cosmological coincidence problems at recombination and at present time and then inspect the third coincidence at the epoch of reionization.
Using current cosmological observations, we then estimate the statistical relevance for this new coincidence through Bayesian model comparison in Sec.~\ref{sec:bayesestimate}.
We test the robustness of our results against effects from changes in priors or reionization history and from allowing free total neutrino mass.
We also estimate the impact of intermediate polarization data from the \emph{Planck} High Frequency Instrument (HFI) and provide an outlook for 21~cm surveys.
We then discuss different interpretations of the results.
In particular, in Sec.~\ref{sec:sfr} we inspect a promising connection between the star-formation history and the cosmological background dynamics.
Finally, we provide conclusions of this work in Sec.~\ref{sec:conclusions}.
Further details of the reionization history adopted and numerical implementations are given in App.~\ref{sec:planckreion}, and a few alternative formulations of the coincidence problems are presented in App.~\ref{sec:furtherviewsofcoincidences}.


\section{A Coincidence at Reionization} \label{sec:coincidence}

\begin{figure*}
 \centering
 \resizebox{0.477\hsize}{!}{\includegraphics{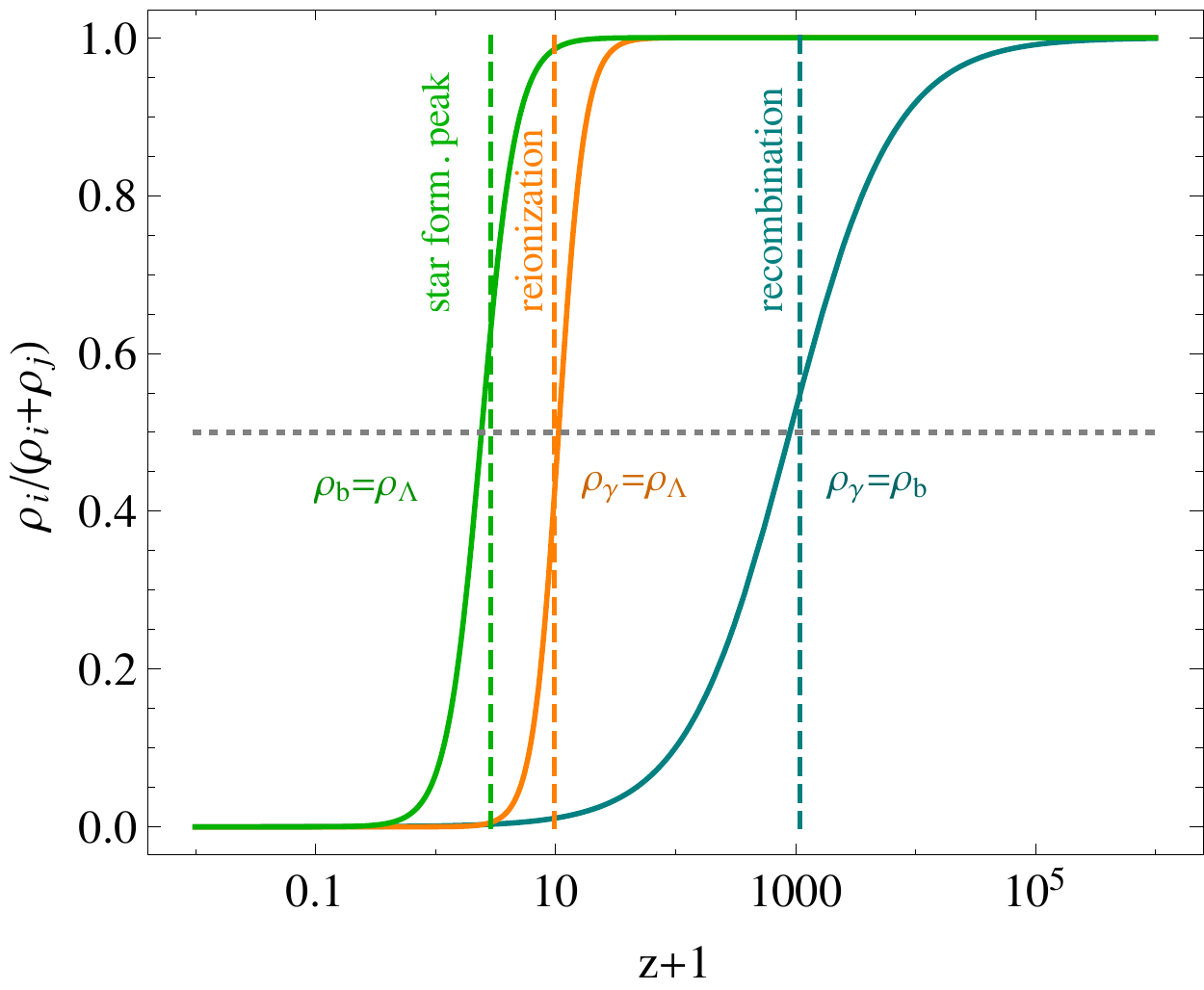}}
 \resizebox{0.493\hsize}{!}{\includegraphics[trim={0mm 0.7mm 0mm 0mm},clip]{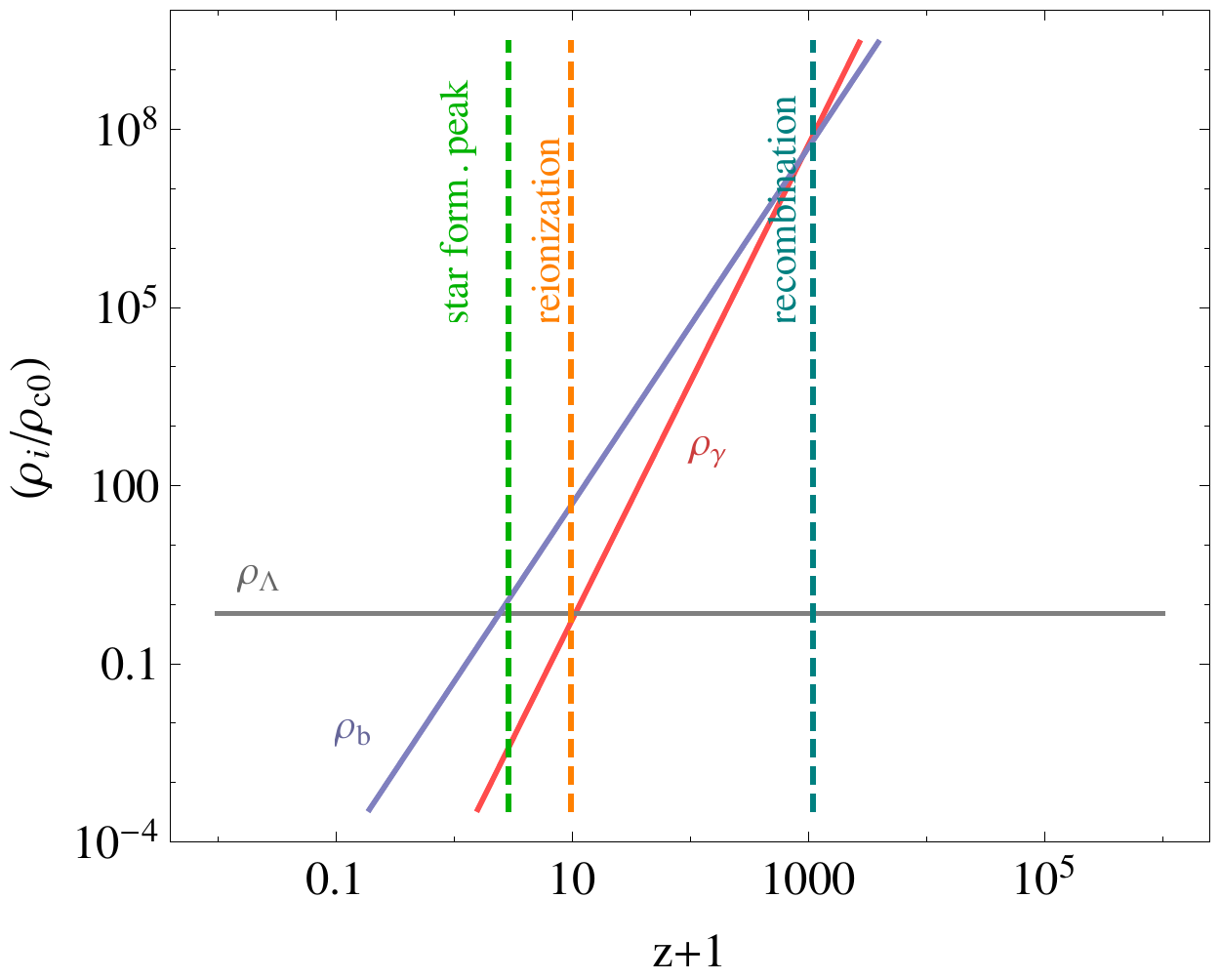}}
 \caption{Cosmological coincidences between equalities in the energy densities of photon radiation $\rho_{\gamma}$, baryonic matter $\rho_{\rm b}$, and the cosmological constant $\rho_{\Lambda}$ with the recombination and reionization epochs and the peak of star formation (representing the \emph{Why Now?} problem) in terms of redshift $z$. 
 The \emph{left panel} shows the fraction of energy densities of one component to the combination with a second component. The \emph{right panel} presents the fraction of the energy densities of the different components to the present critical value $\rho_{{\rm c}0}$.
 The redshift window for equality between $\rho_{\gamma}$ and $\rho_{\Lambda}$ is narrower than those of the traditional cosmological coincidence problems.
 }
 \label{fig:coincidences}
\end{figure*}

Before addressing the coincidence between the equality in the energy densities of CMB radiation and the cosmological constant and the epoch of reionization, we briefly discuss some of the well-known cosmological coincidence problems.
For illustrations, we adopt the Planck cosmological parameters~\cite{Ade:2015xua}: $\Omega_{\rm m}=0.308$ and $\Omega_{\rm b}=0.0484$ for the energy densities of total and baryonic matter, where $\Omega_i \equiv 8\pi G \rho_i(z=0)/(3H_0^2)$ with energy densities $\rho_i$, gravitational constant $G$, Hubble constant $H_0=67.8~{\rm km}\,{\rm s}^{-1}{\rm Mpc}^{-1}$, and speed of light in vacuum set to unity throughout the article. 
We furthermore have $\Omega_{\gamma}=5.38\times10^{-5}$ for the photon radiation.

Probably the most prominent among the cosmic coincidence problems is typically stated as the question of why the energy densities of the cosmological constant $\rho_{\Lambda}$ and the total matter $\rho_{\rm m}$ are comparable today ($z=0$).
The present being observer dependent, one may choose to interpret this coincidence in the context of the existence of the observer and hence relate it to the process of star formation~\cite{Bousso:2008bu} or the production of terrestrial planets~\cite{Lineweaver:2007sla,Lineweaver:2007sla,Loeb:2016vps}.
Motivated by such considerations, we rephrase the problem as the coincidence that the baryonic energy density $\rho_{\rm b}$ becomes comparable to $\rho_{\Lambda}$ around the peak of star formation.
More specifically, $\left.z\right|_{\rho_{\rm b}=\rho_{\Lambda}} = (\Omega_{\rm \Lambda}/\Omega_{\rm b})^{1/3}-1 \simeq 1.4$, which is comparable with the star formation peak $z_{\rm sfp} \approx 1.9$ or with the redshift by which half of the presently observed stellar mass had formed $z=1.3$~\cite{Madau:2014bja}.
Importantly, note that we have interlaced here a further cosmological coincidence, namely the comparable sizes in the energy densities of the baryons and the cold dark matter $\rho_{\rm b}\sim\rho_{\rm c}\sim\rho_{\rm m}$.
Hence, one may interchange the energy densities to restate the problem.
We discuss other combinations in Sec.~\ref{sec:discussion} (also see App.~\ref{sec:furtherviewsofcoincidences}).
In particular, the equality between $\rho_{\rm c}$ and $\rho_{\Lambda}$ at $z=0.4$ matches more closely the age of the Sun and the Earth ($z\approx0.5$), from which one may wish to draw anthropic or multiverse arguments~\cite{Peacock:1999ye,Bousso:2008bu,Lineweaver:2007sla}.
The choice of the baryonic component seems more natural when in the following drawing comparisons to the recombination epoch and the peak of star formation is arguably a less biased, universal and prominent event.

Another well-known coincidence in the cosmic history is associated with the epoch of recombination of neutral hydrogen, briefly described in Sec.~\ref{sec:intro}.
This first of two universal phase transitions of the gas coincides with the equality between the baryonic and photon radiation energy densities.
Recombination occurs when thermal photons no longer ionize neutral hydrogen, whereas the matter-radiation equality can be related to the relative number densities $n_i$.
Hence, the problem can be stated as the coincidence that $n_{\rm b}/n_{\gamma} \sim \alpha^2 m_e/m_p$~\cite{Peacock:1999ye}, where $m_e$ and $m_p$ are the electron and proton rest masses and $\alpha$ is the fine structure constant.
This concurrence of the two epochs may be attributed to a physical process causing a baryon-antibaryon asymmetry of $\sim m_p/(m_e\alpha^2)$, which however begs for an explanation itself.
In specific, equality between the energy densities of photon radiation and baryons occurs at $\left.z\right|_{\rho_{\rm b}=\rho_{\gamma}} = \Omega_{\rm b}/\Omega_{\gamma}-1 \simeq 899$, which is comparable to the recombination redshift $z_{\rm rec}$, approximately the redshift of photon decoupling $z_*\simeq1090$~\cite{Ade:2015xua}.
However, similarly to the late-time coincidence, the correlations can also be interpreted in terms of different matter and radiation components since $\rho_{\rm b}\sim\rho_{\rm c}$ and $\rho_{\gamma}\sim\rho_{\nu}\sim\rho_{\rm rad}$, where the total radiation component $\rho_{\rm rad}$ also includes relativistic neutrinos $\rho_{\nu}$.

Finally, while we have seen that $\rho_{\rm b}\sim\rho_{\Lambda}$ coincides with the peak of star formation and $\rho_{\rm b}\sim\rho_{\gamma}$ with recombination, we find here that the third combination $\rho_{\gamma}\sim\rho_{\Lambda}$ coincides with the second phase transition in the history of our Universe, namely the epoch of reionization.
More specifically, the two energy densities become equal when $\left.z\right|_{\rho_{\Lambda}=\rho_{\gamma}} = (\Omega_{\Lambda}/\Omega_{\gamma})^{1/4}-1 \simeq 9.6$, which compares with the reionization redshift $z_{\rm rei}=8.8$~\cite{Madau:2014bja} (Sec.~\ref{sec:parameters}).
To our knowledge this coincidence has not been pointed out before, which motivates an investigation of its statistical relevance (Sec.~\ref{sec:bayesestimate}).
We note that among the cosmological coincidences discussed here, the coincidence around reionization has the
narrowest redshift window allowing for a correlation.
Inversely, a large range of energy densities $\rho_{\Lambda}$ fall within that window.
This is due to the starkest difference among the powers of the redshift dependence in the energy densities involved (see Fig.~\ref{fig:coincidences}).
As with the other coincidence problems, it should be noted that this concurrence at reionization may also be reinterpreted in terms of different radiation components, for instance the total radiation or relativistic neutrinos.

In summary, we observe three cosmological coincidences between three energy densities at $z\sim\mathcal{O}(10^0)$, $\mathcal{O}(10^3)$, and $\mathcal{O}(10^1)$:
\begin{itemize}
 \item[(i)] $\rho_{\Lambda} \sim \rho_{\rm b}$ at the peak of star formation $z_{\rm sfp}\approx1.9$,
 \item[(ii)] $\rho_{\rm b} \sim \rho_{\gamma}$ at recombination $z_{\rm rec} \approx z_*=1090$,
 \item[(iii)] $\rho_{\gamma} \sim \rho_{\Lambda}$ at the epoch of reionization $z_{\rm rei}\approx8.8$.
\end{itemize}
In terms of the age of the Universe at these redshifts, this corresponds to $\mathcal{O}(10^9)$, $\mathcal{O}(10^5)$, and $\mathcal{O}(10^8)$ years, respectively.
An illustration of the three coincidences is provided in Fig.~\ref{fig:coincidences}.

Finally, while the set of coincidence problems is not uniquely defined, we make an interesting observation for the particular choice made here in that these correlations all seem to revolve around physical processes involving hydrogen.
We refer the reader to Sec.~\ref{sec:discussion} for further discussion and comparison of these coincidences.


\section{A Bayesian Estimate of the Problem} \label{sec:bayesestimate}

In order to provide an estimate for the statistical relevance of the coincidence between the energy densities of photon radiation and the cosmological constant at the epoch of reionization, we resort to Bayesian model comparison.
Hereby, the idea is to assess the significance of this new coincidence by postulating an astrophysical process by which the coincidence naturally manifests itself.
In consequence, this reduces the parameter space by one dimension.
While this can be seen as an \emph{a posteriori} fixing or removal of a parameter, it is \emph{a priori} not guaranteed to be statistically favored.
This is as we are not fixing the base parameter value directly but infer it from the equality of the energy densities.
As such the reduction of parameter volume depends on the narrowness of the window for a coincidence.
Moreover, the model comparison provides an assessment of the maximal gain that could be achieved with a hypothetical parameter-free astrophysical explanation for the concurrence and as such provides an estimate for the relevance of the problem.
We discuss some caveats to this approach in Sec.~\ref{sec:discussion}.
It is worth emphasizing however that the parameter dimension removed in Sec.~\ref{sec:parameters} is the free Thomson scattering optical depth, which is arguably not a fundamentally free parameter (also see Sec.~\ref{sec:sfr}).

In Sec.~\ref{sec:bayesfactor}, we briefly review the model selection we perform based on the Bayes factor.
We discuss the choice of cosmological parameters and priors in Sec.~\ref{sec:parameters}.
In Sec.~\ref{sec:datasets}, we provide details on the cosmological observations adopted for this analysis, summarizing the results in Sec.~\ref{sec:results}.
We check the robustness of these results against changes in priors or reionization history in Sec.~\ref{sec:robustness}, where we also estimate the effects of intermediate \emph{Planck} HFI polarization data and future 21~cm observations.
Finally, we provide a discussion of our findings in Sec.~\ref{sec:discussion}.

\subsection{Bayes factor}\label{sec:bayesfactor}

Let $\theta$ denote a set of parameters of a model $M$ and $D$ be the data.
Bayes' theorem states that
\begin{equation}
 P(\theta | D, M) = \frac{P(D | \theta, M)P(\theta |M)}{P(D | M)}  \,,
 \label{eq:posterior}
\end{equation}
where $P(\theta | D, M)$ is the posterior (or conditional) probability distribution of $\theta$ given the data,
$P(D | \theta, M)$ is the likelihood of the data,
$P(\theta |M)$ is the prior from present knowledge or assumptions about the parameters, and $P(D |M)$ is the evidence, the probability of observing the data for a model (see, e.g.,~\cite{Heavens:2017hkr}).

The Bayesian evidence normalizes the posterior and hence is the integrated numerator of the right-hand-side of Eq.~\eqref{eq:posterior},
\begin{equation}
 P(D | M) = \int d\theta \, P(D | \theta, M) \, P(\theta | M) \,.
 \label{eq:evidence}
\end{equation}
It can be adopted for model selection with the advantage that a higher number of model parameters, which typically yield equal or higher likelihood, is penalized for larger prior volume (Occam's razor), thus quantifying the explanatory power of a model.
The comparison is done by computation of the Bayes factor between the evidences of two models $M_1$ and $M_2$,
\begin{equation}
 B \equiv \frac{P(D | M_1)}{P(D | M_2)} \,.
 \label{eq:bayesfactor}
\end{equation}
Bayes' theorem then implies that $P(M_1|D)/P(M_2|D)=B\,P(M_1)/P(M_2)$, where usually equal priors are assumed.

To then characterize the strength of preference of one model over the other, we adopt the Jeffreys scale, where $B\gtrsim3,10,30,100$ attributes \emph{substantial}, \emph{strong}, \emph{very strong}, and \emph{decisive} support toward $M_1$ over $M_2$, respectively.
The inverse holds for opposite support.

\subsection{Parameters and priors} \label{sec:parameters}

As in the \emph{Planck} analysis~\cite{Ade:2015xua}, we adopt the usual six high-redshift parameters for our base cosmology: the physical baryon and cold dark matter density $\Omega_{\rm b}h^2$, $\Omega_{\rm c}h^2$, the ratio of sound horizon to angular diameter distance at
recombination $\theta_{\rm MC}$, the optical depth to reionization $\tau$,
the scalar tilt $n_s$ and amplitude $A_s$ at $k_∗ = 0.002~\textrm{Mpc}^{-1}$.
Note that unlike the other parameters, $\tau$ acts as a nuisance parameter for the lack of precise astrophysical understanding of the process of reionization.
As such it may be considered not a fundamentally free cosmological parameter and may become specified once reionization is well understood.

With the adoption of the coincidence model, we make the assumption that there is an underlying physical connection of the reionization history with the equality in the energy densities of photon radiation and the cosmological constant that determines the optical depth.
More specifically, requiring equality in the energy densities implies
\begin{equation}
 z_{\rm rei} \overset{!}{=} \left.z\right|_{\rho_{\Lambda}=\rho_{\gamma}} = \left(\frac{\Omega_{\Lambda} h^2}{2.473\times10^{-5}}\right)^{1/4}-1 \,,
 \label{eq:zcoinc}
\end{equation}
where the denominator in Eq.~\eqref{eq:zcoinc} corresponds to the current black body photon radiation energy density parameter $\Omega_{\gamma}h^2$ for $T_{\rm CMB}=2.7255$~\cite{Ade:2015xua}.
The reionization redshift $z_{\rm rei}$ is defined as the redshift at which the reionization fraction reaches half of the maximum~\cite{Lewis:2008wr}.
We adopt the standard reionization history used in the \emph{Planck} analysis~\cite{Ade:2015xua,Lewis:2008wr,camb_notes} (see App.~\ref{sec:planckreion} for more details and Sec.~\ref{sec:robustness} for a discussion of changes in the reionization history).
This produces approximately the same $\tau$ as that of an instantaneous reionization at $z_{\rm rei}$~\cite{Lewis:2008wr}.
Hence, in the concurrent scenario, Eq.~\eqref{eq:zcoinc} eliminates the need to consider the optical depth $\tau$ as a nuisance parameter, restricting the cosmological parameter space to five dimensions.

Importantly, a model where the equality in the energy densities is physical does thus not necessarily have to be considered a beyond-$\Lambda$CDM model, although it could, of course, also be a manifestation of new physics.
We provide a discussion of different interpretations in Sec.~\ref{sec:interpretation}.
In particular, in Sec.~\ref{sec:sfr} we give an example of a possible dynamical origin in concordance cosmology that is connected to the star-formation rate.
To prevent the misinterpretation of our exercise as proposal for a properly alternative cosmology, we will therefore also refer to the six-parameter \emph{vanilla} $\Lambda$CDM model as $\Lambda$CDM+$\tau$, whereas referring to the five-parameter concurrence model as $\Lambda$CDM/Concur.

In addition to our baseline cosmology, we also conduct the model comparison with free total neutrino mass $\sum m_{\nu}$ for three species of degenerate massive neutrinos.
We make no assumption about the mass hierarchy. As in the \emph{Planck} analysis~\cite{Ade:2015xua}, we require positive mass but do not adopt a lower non-vanishing prior, where, however, $\sum m_{\nu}=0.06~\textrm{eV}$ of normal hierarchy is used for the base cosmology.

Finally, in addition to the cosmological parameters, we sample over a range of \emph{Planck} nuisance parameters, following the analysis in Ref.~\cite{Ade:2015xua}.
When computing the Bayes factor, we consider the full set of both cosmological and nuisance parameters as well as the cosmological parameter space marginalized over the nuisance parameters 
(see Table~\ref{tab:results}).

We test both of our models with the same flat priors as were used in the \textit{Planck} analysis~\cite{Ade:2015xua}: $\Omega_{\rm b}h^2\in(0.005,0.1)$, $\Omega_{\rm c}h^2\in(0.001,0.99)$, $100\theta_{\rm MC}\in(0.5,10)$, $\tau\in(0.01,0.8)$,
$n_s\in(0.8,1.2)$, $\ln(10^{10}A_s)\in(2,4)$, and $\sum m_{\nu}\in(0,5)$.
We refer to Sec.~\ref{sec:robustness} for a discussion of the effect of changing prior ranges and to Ref.~\cite{Ade:2015xua} for details on \emph{Planck} nuisance parameter priors.

\subsection{Cosmological datasets} \label{sec:datasets}

We adopt a range of geometric probes to constrain the cosmological background parameters, including data from supernovae (SNe), baryon acoustic oscillations (BAOs), the local Hubble expansion, and distance information from the CMB acoustic peaks.
More specifically, for the SN~Ia luminosity distances, we use the Joint Lightcurve Analysis (JLA)~\cite{Betoule:2014} dataset with records from the Sloan Digital Sky Survey (SDSS) plus the C11 compilation~\cite{Conley:2011} that includes supernovae from the Supernovae Legacy Survey~\cite{Sullivan:2011}, the Hubble Space Telescope (HST)~\cite{Suzuki:2012}, and several nearby experiments, consisting of 740 SNe Ia.
For the Hubble constant, we adopt the measurement of Ref.~\cite{Riess:2011yx}, $H_0=73.8 \pm 2.4$~km~ $\rm{s}^{-1}$~$\rm{Mpc}^{-1}$, obtained from optical and infrared observations of $\sim600$ Cepheids in host galaxies of 8 SNe~Ia with the HST Wide Field Camera 3.
BAO information is included from the 6dF Galaxy Redshift Survey (6dFGRS) at $z_{\rm eff}=0.106$~\cite{Beutler:2011}, SDSS DR7 at $z_{\rm eff}=0.15$~\cite{Ross:2014}, and the Baryon Oscillation Spectroscopic Survey (BOSS) DR11 at $z_{\rm eff}=0.57$~\cite{Anderson:2013}.

Finally, we use CMB temperature, polarization, and lensing data from \emph{Planck} 2015~\cite{Adam:2015}.
More specifically, we utilize the high-$\ell$ $TT$ likelihood ($\ell = 30-2508$), the low-$\ell$ $TEB$ dataset ($\ell=2-29$) for the $TT$, $EE$, $BB$, and $TE$ joint likelihood (see Sec.~\ref{sec:futuredata} for estimates with intermediate HFI data); and the $T+P$ baseline lensing likelihood with both $T$ and $P$ SMICA reconstruction.
The CMB temperature anisotropy power spectrum is sensitive to the combination $A_s e^{-2\tau}$.
This leaves a degeneracy between the two parameters, which can however be alleviated with the inclusion of the CMB lensing data~\cite{Ade:2015xua}.
The low-multipole reionization feature in the $E$-mode polarization power spectrum instead scales as $\sim\tau^2$.
A $\tau^2$ dependence can also enter the $B$-mode polarization due to primordial tensor fluctuations scattering off reionized matter.
These data hence constrain the viability range of $\tau$ values, which are used together with background constraints to infer bounds on $z_{\rm rei}$ (see App.~\ref{sec:planckreion}).

\subsection{Results} \label{sec:results}

We use the publicly available \textsc{CosmoMC}~\cite{Lewis:2013hha,Lewis:2002ah} package (November 2016 version) to produce Markov Chain Monte Carlo (MCMC) samples.
The code employs the Metropolis-Hastings algorithm~\cite{Metropolis:1953am,Hastings:1970aa} for sampling and Bayesian parameter inference with Gelman-Rubin statistic $R$~\cite{Gelman:1992zz} for convergence testing, where we require $R-1<10^{-3}$ for our runs in eight parallel chains. This generally produces $\mathcal{O}(10^6)$ samples in parameter space.
For the computation of the CMB anisotropies, we use the embedded \textsc{CAMB}~\cite{Lewis:1999bs} Einstein-Boltzmann linear theory solver.

We then use \textsc{MCEvidence}~\cite{Heavens:2017hkr,Heavens:2017afc} to compute the evidences from the collection of these chains with and without marginalization over all nuisance parameters (17 co-dimensions).
\textsc{MCEvidence} is designed to compute the Bayesian evidence from MCMC sampled posterior distributions.
This is generally difficult to do with MCMC samples due to the required normalization in Eq.~\eqref{eq:posterior}.
The code employs $k$-th nearest-neighbor distances in parameter space based on the Mahalanobis metric.
We shall quote results for $k=1$, which was found to be the most accurate choice~\cite{Heavens:2017hkr,Heavens:2017afc}.
To reduce parameter correlations, one may wish to thin the chains.
The effect on the Bayes factor, however, is small as was shown in Ref.~\cite{Heavens:2017hkr}, finding limited impact even for aggressive thinning.

\begin{figure}
   \resizebox{0.554\hsize}{!}{\includegraphics[trim={2mm 2mm 2.3mm 2.3mm},clip]{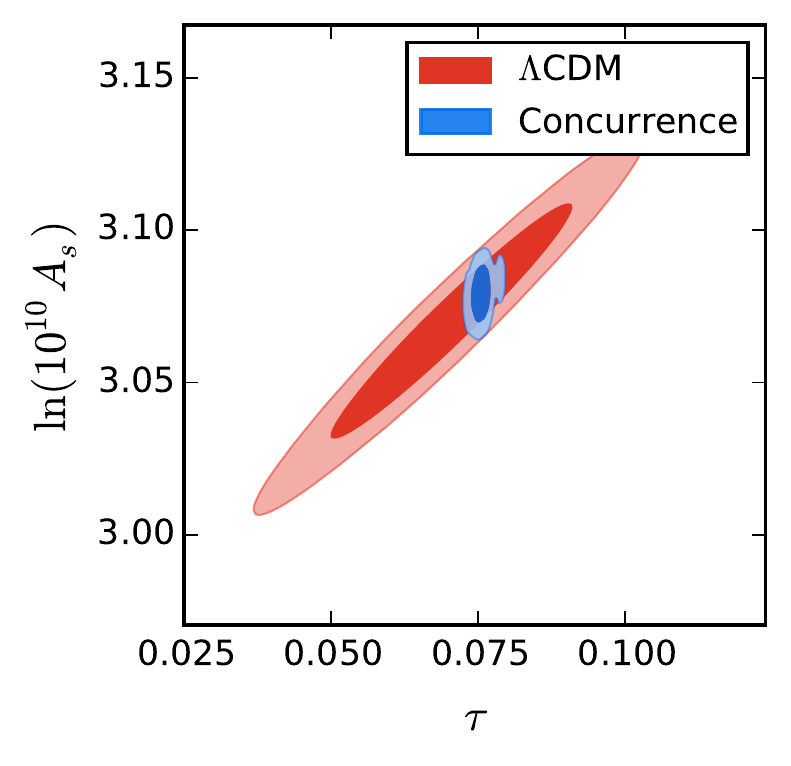}}
   \resizebox{0.434\hsize}{!}{\includegraphics[trim={18.5mm 2.45mm 2.3mm 1.0mm},clip]{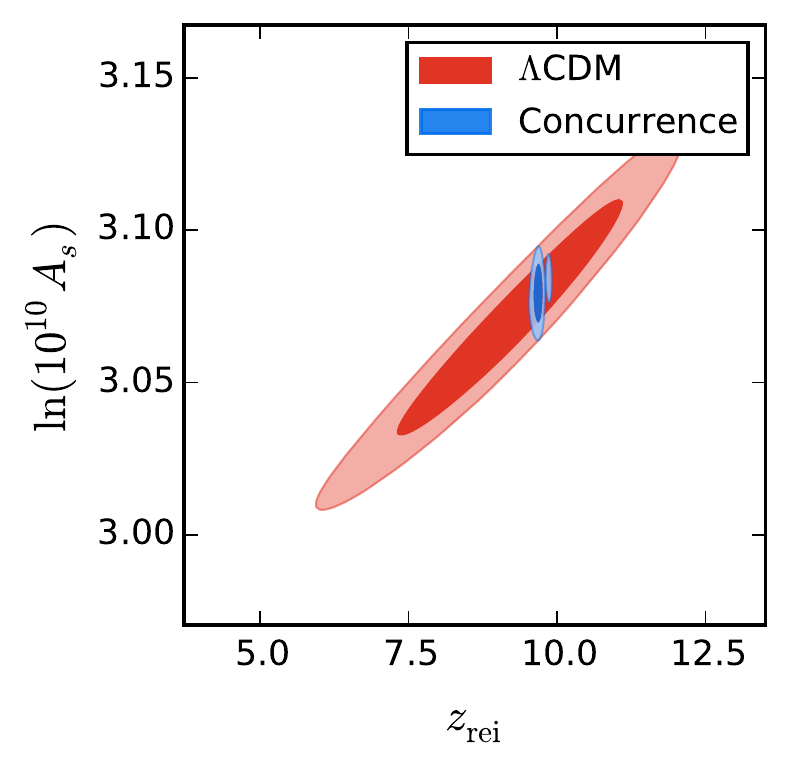}}
 \caption{Contours of 2D-marginalized 68\% and 95\% confidence boundaries for $\Lambda$CDM+$\tau$ and $\Lambda$CDM/Concur(rence) obtained from current cosmological observations (Sec.~\ref{sec:datasets}). The allowed parameter range of the amplitude of primordial fluctuations $A_{\rm s}$ against the Thomson scattering optical depth $\tau$ (\emph{left panel}) and the reionization redshift $z_{\rm rei}$ (\emph{right panel}) is strongly reduced when assuming Eq.~\eqref{eq:zcoinc}.}
 \label{fig:contours}
\end{figure}

From the evidences obtained in this process, we compute the Bayes factors in Eq.~\eqref{eq:bayesfactor} between \emph{vanilla} $\Lambda$CDM ($\Lambda$CDM+$\tau$) and the concurrence model ($\Lambda$CDM/Concur), which are presented in Table~\ref{tab:results}.
These results are denoted by $B_{\rm current}$ to distinguish the use of current cosmological data (Sec.~\ref{sec:datasets}) from the forecasts used in Sec.~\ref{sec:futuredata}.
The classification as \emph{Planck} standard further distinguishes them from other tests conducted in Sec.~\ref{sec:robustness}.
The marginalized parameter constraints and best-fit values for the base parameters and selected derived parameters of $\Lambda$CDM+$\tau$ and $\Lambda$CDM/Concur are summarized in Table~\ref{tab:constraints}.
To illustrate the reduction in posterior parameter distribution, we show the 2D-marginalized contours for the amplitude of the initial power spectrum $\ln(10^{10}A_s)$ against the optical depth $\tau$ and reionization redshift $z_{\rm rei}$ in Fig.~\ref{fig:contours}.

For the runs in standard configuration, we find \emph{very strong} preference on the Jeffreys scale for $\Lambda$CDM/Concur over $\Lambda$CDM+$\tau$, regardless of marginalization over nuisance parameters.
Hence, a hypothetical parameter-free physical connection supporting an exact match in reionization redshift to the redshift of equality in the cosmological constant and photon radiation energy densities would yield a high explanatory gain for current cosmological observations.
These first results support the view that the coincidence at the reionization epoch may be characterized as a problem.

In the following, we will analyze the robustness of these findings against changes in the configuration of the MCMC runs and possible new observations.

\begin{table*}
  \begin{tabular}{l|c|c|c|c|c}
   \hline\hline
   Configuration & Dimensions & $B_{\rm current}$  & $B_{\rm forecast\:1}$  & $B_{\rm forecast\:2}$  & $B_{\rm forecast\:3}$ \\
   \hline
   \emph{Planck} standard & $6 (-1)$ & 36 & 9 & 1/5 & 408 \\
    & $23 (-1)$ & 34 & 8 & 1/4  & 362 \\
   Tighter $\tau$ prior ($\times[3,\frac{1}{3}]$) & $6 (-1)$ & $39$ & -- & -- & -- \\
    &  $23 (-1)$ & $33$ & -- & -- & -- \\
   Reionization width ($\times3$) & $6 (-1)$ & $41$ & -- & -- & -- \\
    & $23 (-1)$ & $31$ & -- & -- & -- \\
   Free $\sum m_{\nu}$ & $7 (-1)$ & $36$ & 9 & 1  & 393 \\
    & $24 (-1)$ & $25$ & 7 & 1  & 387 \\
   \hline\hline
  \end{tabular}
 \caption{Bayes factors between $\Lambda$CDM/Concur(rence) and concordance $\Lambda$CDM+$\tau$ for current cosmological data ($B_{\rm current}$), estimated intermediate \emph{Planck} HFI $EE$ data ($B_{\rm forecast\:1}$), and future 21~cm surveys ($B_{\rm forecast\:2}$ and $B_{\rm forecast\:3}$ for centering around mean of $\Lambda$CDM+$\tau$ and $\Lambda$CDM/Concur, respectively).
Computations are done both by including the Planck nuisance parameters and by marginalizing over them, which corresponds to 23 and 6 parameter space dimensions for the standard configuration.
The concurrent model is reduced by one dimension with the identification of the reionization redshift with the redshift of equality between the energy densities of photon radiation and the cosmological constant.
We adopt the \emph{Planck} prior on $\tau$ and also analyze the impact on the Bayes factors if tightening the lower and upper bounds of the prior by a factor of three and $1/3$, respectively.
Furthermore, we compute the Bayes factors for increased redshift width in the reionization history, and when allowing for a free sum of neutrino masses.
Given the small effect of changes in the $\tau$ prior and the reionization history, we omit the forecasts in those scenarios.
Details are provided in Sec.~\ref{sec:bayesestimate}.
}
 \label{tab:results}
\end{table*}

\begin{table*}
  \begin{tabular}{l|cc|cc|cc|cc}
   \hline\hline
   Parameter & \multicolumn{2}{c|}{$\Lambda$CDM+$\tau$} & \multicolumn{2}{c|}{$\Lambda$CDM/Concur} & \multicolumn{2}{c|}{$\Lambda$CDM+$\tau$+$\nu$} & \multicolumn{2}{c}{$\Lambda$CDM/Concur+$\nu$} \\
   \hline
   $\Omega_{\rm b}h^2$ & $0.02232\pm0.00019$ & 0.02223 & $0.02234\pm0.00020$ & 0.02221
                       & $0.02234\pm0.00020$ & 0.02228 & $0.02234\pm0.00019$ & 0.02227 \\
   $\Omega_{\rm c}h^2$ & $0.1178\pm0.0012$ & 0.1172 & $0.1177\pm0.0010$ & 0.1170
                       & $0.1176\pm0.0013$ & 0.1180 & $0.1176\pm0.0010$ & 0.1175 \\
   $100\theta_{\rm MC}$ & $1.04113\pm0.00041$ & 1.04111 & $1.04117\pm0.00040$ & 1.04088
                        & $1.04115\pm0.00041$ & 1.04112 & $1.04118\pm0.00040$ & 1.04117\\
   $\tau$ & $0.070\pm0.013$ & 0.077 & -- & --
          & $0.075\pm0.016$ & 0.061 & -- & -- \\
   $\ln(10^{10}A_s)$ & $3.070\pm0.024$ & 3.082 & $3.0793\pm0.0057$ & 3.0761
          & $3.078\pm0.030$ & 3.046 & $3.0792\pm0.0057$ & 3.0772 \\
   $n_s$ & $0.9693\pm0.0043$ & 0.9725 & $0.9702\pm0.0039$ & 0.9715
         & $0.9697\pm0.0046$ & 0.9660 & $0.9698\pm0.0038$ & 0.9708 \\
   $\sum m_{\nu}$ & -- & -- & -- & --
         & $<0.21$ & 0.01 & $<0.17$ & 0.06 \\
   \hline
   $\tau$ & -- & -- & $0.0755\pm0.0012$ & 0.0754
          & -- & -- & $0.0753\pm0.0013$ & 0.0754 \\
   $z_{\rm rei}$ & $9.2\pm1.2$ & 9.9 & $9.689\pm0.060$ & 9.697
                 & $9.6\pm1.4$ & 8.3 & $9.672\pm0.076$ & 9.694 \\
   $H_0$ & $68.12\pm0.53$ & 68.31 & $68.24\pm0.47$ & 68.26
         & $68.01\pm0.64$ & 68.46 & $68.08\pm0.62$ & 68.19 \\
   $\Omega_{\rm m}$ & $0.3037\pm0.0069$ & 0.3001 & $0.3021\pm0.0059$ & 0.3003
                    & $0.3047\pm0.0078$ & 0.2997 & $0.3039\pm0.0076$ & 0.3021 \\
   $\sigma_8$ & $0.8164\pm0.0092$ & 0.8199 & $0.8194\pm0.0040$ & 0.8166
              & $0.812\pm0.014$ & 0.817 & $0.813\pm0.014$ & 0.818 \\
   \hline\hline
  \end{tabular}
  \caption{Mean, standard deviation, and best-fit values for the free cosmological parameters and for a selection of derived parameters. For neutrinos, the upper 95\% confidence level is quoted on the total mass. We do not provide constraints on the \emph{Planck} nuisance parameters (see Ref.~\cite{Ade:2015xua} for more details).}
  \label{tab:constraints}
\end{table*}

\subsection{Robustness and future data} \label{sec:robustness}

While we find an interesting statistical preference for an equality of the energy densities of photon radiation and cosmological constant at reionization, it is not clear how easily these findings can be manipulated.
We therefore check the robustness of our results presented in Sec.~\ref{sec:results} against changes in the prior of optical depth (Sec.~\ref{sec:tauprior}), the reionization history (Sec.~\ref{sec:reionizationhistory}), or total neutrino mass (Sec.~\ref{sec:freeneutrinomass}).
Furthermore, we estimate the impact that future data may have on the reported Bayes factors in Sec.~\ref{sec:futuredata}.

\subsubsection{Changes in prior of optical depth} \label{sec:tauprior}

As we discussed in Sec.~\ref{sec:bayesfactor} and can be seen in detail in Eqs.~\eqref{eq:evidence} and \eqref{eq:bayesfactor}, the Bayes factors in Table~\ref{tab:results} carry a dependence on the priors adopted in the analysis.
We check the robustness of our results obtained in Sec.~\ref{sec:results} with a change of the prior on the optical depth $\tau$.
For this purpose, we increase and decrease the lower and upper bounds of the flat prior by a factor of 3 and 1/3, respectively.
More specifically, we change the prior in Sec.~\ref{sec:parameters} from $\tau\in(0.01,0.8)$ to $\tau\in(0.03,0.27)$ such that it still lies outside the $3\sigma$ region.

The Bayes factors produced by this analysis are presented in Table~\ref{tab:results}.
The preference for a coincidence remains \emph{very strong}, confirming the robustness against the change of the $\tau$ prior.
This does, however, not preclude that the results are not affected by more drastic changes in the prior bounds or the assumption of flatness.
It is worth noting, however, that Ref.~\cite{Heavens:2017hkr} analyzed a range of extra-parameter models with Bayes factors of up to half of the strength of the results reported in Table~\ref{tab:results}, finding a requirement to reduce priors by about a factor of 20 to turn the preferences around.

\subsubsection{Redshift width of reionization history} \label{sec:reionizationhistory}

To produce our results in Sec.~\ref{sec:results}, we have adopted the same reionization history as that of the \emph{Planck} analysis (see App.~\ref{sec:planckreion}).
As pointed out in Sec.~\ref{sec:parameters}, the reionization redshift, defined at half of the maximal reionization fraction, produces approximately the same $\tau$ as that of an instantaneous reionization at $z_{\rm rei}$ if reionization occurs in the matter-dominated regime.
The CMB is only affected by $\tau$ (see Sec.~\ref{sec:datasets}), rather than directly through the reionization history and its
redshift evolution.
As such, our results presented in Sec.~\ref{sec:results} are largely insensitive to changes in this evolution.
To reinforce this point, we redo the analysis with increased redshift width $\Delta_z$ in the reionization history.
More specifically, we enhance $\Delta_z$ by a factor of 3 to $\Delta_z=1.5$.

We present the Bayes factors of this analysis in Table~\ref{tab:results}.
With the preference for a coincidence remaining \emph{very strong}, we confirm the robustness against the change of reionization history.
Note however that the result may change if considering more exotic reionization models (see, e.g.,~\cite{Heinrich:2016ojb}) and we leave a more detailed analysis of their effects to future work.

\subsubsection{Free neutrino mass} \label{sec:freeneutrinomass}

We furthermore test for changes in the Bayes factors of Table~\ref{tab:results} when allowing for a free sum of neutrino masses $\sum m_{\nu}$.
As detailed in Sec.~\ref{sec:parameters}, this implies rerunning the chains for $\sum m_{\nu}\in(0,5)$ instead of $\sum m_{\nu} = 0.06$~eV, where we drop units for convenience.

Since allowing for this variation in neutrino mass can be considered the standard cosmology up to a mass hierarchy dependent lower bound, besides presenting the Bayes factors of this analysis in Table~\ref{tab:results}, we also provide the resulting marginalized parameter constraints in Table~\ref{tab:constraints}.
Interestingly, we find that when not marginalizing over the \emph{Planck} nuisance parameters in the computation of the Bayes factor, the preference for the coincidence lowers from \emph{very strong} to \emph{strong}.
This reflects a level of degeneracy between the effect of neutrino mass sum $\sum m_{\nu}$ and the optical depth $\tau$, and hence the reionization redshift $z_{\rm rei}$.
One can also notice this degeneracy by the effect on the marginalized constraints on $\ln(10^{10}A_s)$ (and $n_s$), which displays a degeneracy with $\tau$ (Fig.~\ref{fig:contours}), and correspondingly with $\sigma_8$, also affecting $H_0$ and $\Omega_{\rm m}$ (see Table~\ref{tab:constraints}).

\subsubsection{Impact estimate of future data} \label{sec:futuredata}

Finally, we also test how the results may change with future data.
In particular, a low value of $\tau$ has been reported from the intermediate analysis of the \emph{Planck} HFI polarization data~\cite{Aghanim:2016yuo,Adam:2016hgk}, who found a constraint of $\tau=0.055\pm0.009$~\cite{Aghanim:2016yuo} from the $E$-mode power spectrum only.
This lower value of $\tau$ (and also of $z_{\rm rei}$) is in better agreement with astrophysical measurements (Sec.~\ref{sec:intro}).
While this value is still in good agreement with the loosely constrained optical depth of $\Lambda$CDM+$\tau$ in Table~\ref{tab:constraints}, it is in slight tension with
the more tightly constrained larger optical depth found for $\Lambda$CDM/Concur.
Importantly, however, one should note that the crucial quantity for the coincidence is the reionization redshift, which still lies well within the $2\sigma$ region of Refs.~\cite{Aghanim:2016yuo,Adam:2016hgk}.
Considering the analysis of similar scenarios with extra degrees of freedom introduced on top of $\Lambda$CDM+$\tau$ that show similar deviations in the marginalized constraints~\cite{Heavens:2017hkr}, we do not expect a difference in the Bayes factor from an inclusion of HFI data that could be strong enough to overturn the preference toward $\Lambda$CDM+$\tau$ instead.
Since the HFI data is not public yet, we estimate the robustness of the Bayes factors in Table~\ref{tab:results} against this data by rerunning the chains with inclusion of the intermediate marginalized $\tau$ constraint.
Because of the weak dependence of the Bayes factor on the tightening of the prior on the optical depth (Sec.~\ref{sec:tauprior}) or the redshift width in the reionization history (Sec.~\ref{sec:reionizationhistory}), however, we only reconsider the standard configuration as well as runs allowing free sum of neutrino masses, where the effect on $B$ was larger (Sec.~\ref{sec:freeneutrinomass}).
We follow the same approach with other impact estimates for future data.

Of particular interest for exploring the epoch of reionization, will be experiments measuring the redshift distribution of neutral hydrogen in the intergalactic medium with its hyperfine 21~cm transition~\cite{Zaroubi:2012in,Liu:2015txa}.
This will provide a direct probe of the reionization history.
Ref.~\cite{Liu:2015txa} estimated a constraint of $\sigma_{\tau} \approx 0.0008$ from 21~cm observations with the Hydrogen Epoch of Reionization Array~\cite{Pober:2013jna} and $\sigma_{\tau} \approx 0.0006$ with the Square Kilometre Array (SKA)~\cite{Mellema:12} in combination with \emph{Planck} data.
For a simple estimation of how future 21~cm measurements can affect the statistical relevance of the coincidence, we compute the Bayes factor from rerunning the chains for the standard and free neutrino mass configurations with inclusion of a $\tau$ constraint centered around the mean of either the baseline $\Lambda$CDM+$\tau$ or the concurrence model runs (see Table~\ref{tab:constraints}) with the expected SKA $\sigma_{\tau}$.

In summary, in addition to the observational datasets in Sec.~\ref{sec:datasets}, we alternately introduce the constraints on optical depth that one may expect from future observations:
\begin{eqnarray}
 \textrm{forecast 1} & : & \tau =  0.055\pm0.009 \ \ \ \ \textrm{(\emph{Planck}~lowE~HFI)} \nonumber\\
 \textrm{forecast 2} & : & \tau =  0.0704\pm0.0006 \ \textrm{(SKA~21~cm; free $\tau$)} \nonumber\\
 \textrm{forecast 3} & : & \tau =  0.0755\pm0.0006 \ \textrm{(SKA~21~cm; concur.)} \nonumber
\end{eqnarray}
We then rerun the chains and recompute the Bayes factors in Table~\ref{tab:results}.
For forecast~1, we use the most stringent of the $\tau$ constraints found in Ref.~\cite{Aghanim:2016yuo}, where the likelihood adopted sampled $\tau$ with fixed best-fit values for $A_s e^{-2\tau}$ and other cosmological parameters.

We find that although the mean of the optical depth for forecast~1 deviates more strongly from the coincidence value, the preference for a coincidence remains \emph{substantial}.
With forecast~2, we find \emph{substantial} evidence against an exact correlation in the case of fixed total neutrino mass whereas for free $\sum m_{\nu}$ neither of the models are preferred over the other.
This is due to a partial compensation of the tension in the coincidence model with a neutrino mass detection.
Finally, forecast~3 can make a \emph{decisive} statement on the preference of the concurrence.
However, one should not expect future constraints to center exactly around the current mean values.
The reported numbers should therefore only be interpreted as a rough guide.
Importantly, the future measurements will dominate the $\tau$ constraint.
Comparably, the data described in Sec.~\ref{sec:datasets} have the effect of changing the 1D-marginalized constraints on the optical depth to $\tau=0.0599\pm0.0075$, $\tau=0.07040\pm0.00062$, and $0.07482\pm0.00072$ for the $\Lambda$CDM+$\tau$ runs with forecast~1, 2, and 3, respectively.

\subsection{Discussion} \label{sec:discussion}

In Sec.~\ref{sec:results}, we have found that a hypothetical parameter-free model in which the concurrent epoch of reionization and the $\rho_{\Lambda}$-$\rho_{\gamma}$
equality naturally manifests itself would significantly increase the explanatory power for current cosmological observations.
From this perspective, the coincidence may therefore be considered of high enough statistical relevance to consider it an interesting problem.
In Sec.~\ref{sec:robustness}, we have further observed that these findings are robust against changes in $\tau$ prior and reionization history, and that future data has the potential to turn the Bayes factor against a concurrence.
The approach we have pursued here, however, can only serve as a guideline and we shall discuss some concerns and caveats to it in Sec.~\ref{sec:concerns}.
In Sec.~\ref{sec:comparison}, we then compare the coincidence at reionization with the traditional coincidence problems to assess whether it should be viewed in the same vein.
Finally, we provide different interpretations for how such a concurrence could arise in Sec.~\ref{sec:interpretation}, focusing in Sec.~\ref{sec:sfr} on an interesting connection of cosmological background dynamics with the star-formation rate.

\subsubsection{Concerns and caveats} \label{sec:concerns}

The strongest caveat to the analysis of the concurrence model is that it lacks a proper physical motivation.
A major concern is therefore that we have constructed a model \emph{a posteriori} purely on inductive grounds.
An inevitable question is thus whether one could not simply pick other equalities in the energy densities and find a suitable coincidental epoch or event in the history of the Universe.
The same concern, however, applies to the traditional cosmic coincidence problems.
With the aim of this paper to assess whether the coincidence at the reionization epoch should be viewed as a comparable problem, 
we shall provide a comparison between the coincidences in Sec.~\ref{sec:comparison} (also see Sec.~\ref{sec:coincidence}).

Another concern is that one may not expect two subdominant energy components to have any influence on the process of reionization that lies well within the matter-dominated regime.
Another view on this aspect, however, is presented if one considers the peak of matter domination, where $\Omega_{\rm m/b}(a) \equiv 8\pi G \rho_{m/b}/(3H^2)$ becomes maximal.
This is given by the expression $z=(3\Omega_{\Lambda}/\Omega_{\rm rad})^{1/4}-1\approx11$ for massless neutrinos, which shows the same dependence on cosmological parameters as Eq.~\eqref{eq:zcoinc}.
We will see in Sec.~\ref{sec:sfr} that such a relation may also emerge from a reionization threshold in the star-formation history.

Finally, as pointed out in Sec.~\ref{sec:parameters}, the optical depth $\tau$ may be viewed as a non-fundamental parameter that will be fixed once the reionization processes are well understood.
As such, one may expect that any parameter-free reionization model that predicts $\tau$ within the range of values allowed by current data should increase the evidence over $\Lambda$CDM+$\tau$.
We therefore emphasize again that the better performance of $\Lambda$CDM/Concur should not be viewed as evidence against $\Lambda$CDM but it also shows that the coincidence at reionization is an interesting problem from the
statistical perspective with high explanatory potential.

\subsubsection{Comparison to traditional coincidence problems} \label{sec:comparison}

As in this paper, we wish to assess whether the reionization coincidence should be viewed as a problem of similar sort to the established cosmological coincidence problems, we briefly inspect the range of redshifts one can obtain from the exchange of radiation components and compare it to the analogous alterations
for the other coincidences.
For that estimation, we adopt the best-fit values from $\Lambda$CDM+$\tau$ (Table~\ref{tab:constraints}).
The equalities from exchanges in the energy density components are then summarized as follows:
\begin{itemize}
 \item[(ii)] $(\rho_{\rm b}, \rho_{\rm c}, \rho_{\rm m}) \sim \rho_{\Lambda}$ covers a redshift range for equalities of $z\in(0.3,1.5)$;
 \item[(ii)] $(\rho_{\rm b}, \rho_{\rm c}, \rho_{\rm m}) \sim (\rho_{\gamma}, \rho_{\nu}, \rho_{\rm rad})$ allows for correlations in a broad redshift range of $z\in(530,8150)$;
 \item[(iii)] $(\rho_{\gamma}, \rho_{\nu}, \rho_{\rm rad}) \sim \rho_{\Lambda}$ covers a narrow redshift range of $z\in(8.4,10.8)$, assuming massless neutrinos.
\end{itemize}
Hence, under the aspect of broadness of
redshift windows for potential correlations,
the coincidence (iii)
at the epoch of reionizaion seems the most remarkable among the three.

It is worth noting that the redshift ranges of (i)-(iii) do not overlap and $z\sim\mathcal{O}(10^0)$, $\mathcal{O}(10^3)$, and $\mathcal{O}(10^1)$, respectively; or in terms of age: $\mathcal{O}(10^9)$, $\mathcal{O}(10^5)$, and $\mathcal{O}(10^8)$ years (Sec.~\ref{sec:coincidence}).
In terms of epochs or events in the history of the Universe that these redshifts could be correlated with, for (i) and (iii) there is nothing of the same bearing as the first and second phase transition of the hydrogen gas, while for (iii) at smaller redshifts, choices are less obvious.
In particular, as pointed out in Sec.~\ref{sec:coincidence}, one may choose the equality between $\rho_{\rm c}$ and $\rho_{\Lambda}$ matching more closely the age of the Solar System.
This combination can be interesting as it addresses the dark sector in conjunction but it also detaches the intersection from the known physical contributions.
One can also consider redshift derivatives of the relative dominance of the matter and $\Lambda$ energy density components instead, e.g., Eq.~\eqref{eq:ccderivativecoinc} that peaks precisely today.
  
Adopting the components $\rho_{\rm b}$, $\rho_{\gamma}$, and $\rho_{\Lambda}$ as in Sec.~\ref{sec:coincidence}, we furthermore compare the relative proximity of the redshifts and cosmic ages at the equalities to those of the different coincidental epochs.
The coincidental redshifts around recombination and the peak of star formation lie within $\sim20\%$ and $\sim25\%$, respectively, whereas there is a $\sim10\%$ agreement around reionization.
In terms of the age of the Universe at those concurrences, we find approximately $30\%$, $25\%$, and $10\%$ agreement.
We note therefore, that the coincidence at reionization is the closest one.
Even if future data will favor a lower redshift, e.g., $z_{\rm rei}=8.0$ (best fit of $\Lambda$CDM+$\tau$ with forecast~1), the agreement remains within about $20\%$ in redshift and $25\%$ in age.
This is also comparable to the typically considered $\rho_{{\rm m}0}$-$\rho_{\Lambda}$ coincidence problem, where the current age is $\sim25\%$ larger than the age at which $\rho_{\rm m}$ equals $\rho_{\Lambda}$.
Similarly, with the redshift of the last-scattering surface $z_*=1089.80\pm0.29$ clearly not agreeing with exact equality between $\rho_{\rm b}$ and $\rho_{\gamma}$, a much more tightly constrained value around a lower reionization redshift, e.g., $z_{\rm rei}\approx8.0$, would thus remain a comparable correlation.

These considerations therefore motivate the concurrence between the epoch of reionization and the time of equality between the energy densities of photon radiation and the cosmological constant to be viewed at a similar level to the traditional cosmological coincidence problems.
This is, even if the current statistical preference of an exact match in redshift will cease with future data.

\subsubsection{Interpretations of a possible concurrence} \label{sec:interpretation}

If one is to view the coincidence at the reionization epoch as a problem worth investigating, what then could be the physical nature of such a concurrence?
We shall briefly inspect three possible but non-exhaustive scenarios.
One may, for instance, want to explore whether
\begin{itemize}
 \item[(i)] the same dynamical relations between the energy densities (or between the fundamental constants involved) govern the physics of reionization or even of two (or of all three) different epochs;
 \item[(ii)] reionization processes produce the cosmological constant or dark energy;
 \item[(iii)] the cosmological constant or dark energy induces reionization at the atomic level.
\end{itemize}
Scenario~(i) seems the most plausible explanation.
It could easily be a phenomenological feature of standard cosmology.
In this respect, it is interesting to observe that correlations with $\rho_{\Lambda}$ occur with reionization in the early star-formation history and with the late-time coincidence at the peak of star formation.
Relations connecting the star-formation peak, reionization, and possibly even recombination with the dynamics of the background densities are therefore well worth investigating.
We provide some considerations in Sec.~\ref{sec:sfr}.
In general, considering relations that are connected among the different epochs, scenario~(i) would provide an explanation for why two subdominant energy density components should correlate with a dominant universal process like reionization without invoking new physics.

Considering scenario~(ii), one should keep in mind that the cosmological constant only contributes dynamically in the late Universe and if it was to turn on at reionization, it would hardly leave an observable signature.
It is intriguing to speculate how early galaxies could induce a dark energy term.
With $\rho_{\Lambda}$ being subdominant, one could imagine a physical process by which star formation gradually increases or decreases the dark energy contribution; a process which saturates around the star formation peak at $z_{\rm sfp}\approx1.9$ and its steep decline thereafter. Given that dark energy only dominates at $z\lesssim0.3$ such a contribution may easily appear like a constant term when it comes to dominate. If deviating strongly from a constant at redshifts $z\gtrsim(1-2)$, this could leave a testable signature.

Finally, it is more difficult to imagine how the cosmological constant or vacuum fluctuations would induce reionization at a particle level for scenario~(iii) as it involves vastly different energy scales than what would be required for reionization and its energy density is subdominant at this epoch.
On the other hand, the search for such a process, possibly involving some exotic dark energy model, may be interesting as an explanation for an additional, early reionization source in case of observing high reionization redshifts.

\subsubsection{A view from the star-formation history} \label{sec:sfr}

\begin{figure*}
 \centering
 \resizebox{0.3345\hsize}{!}{\includegraphics{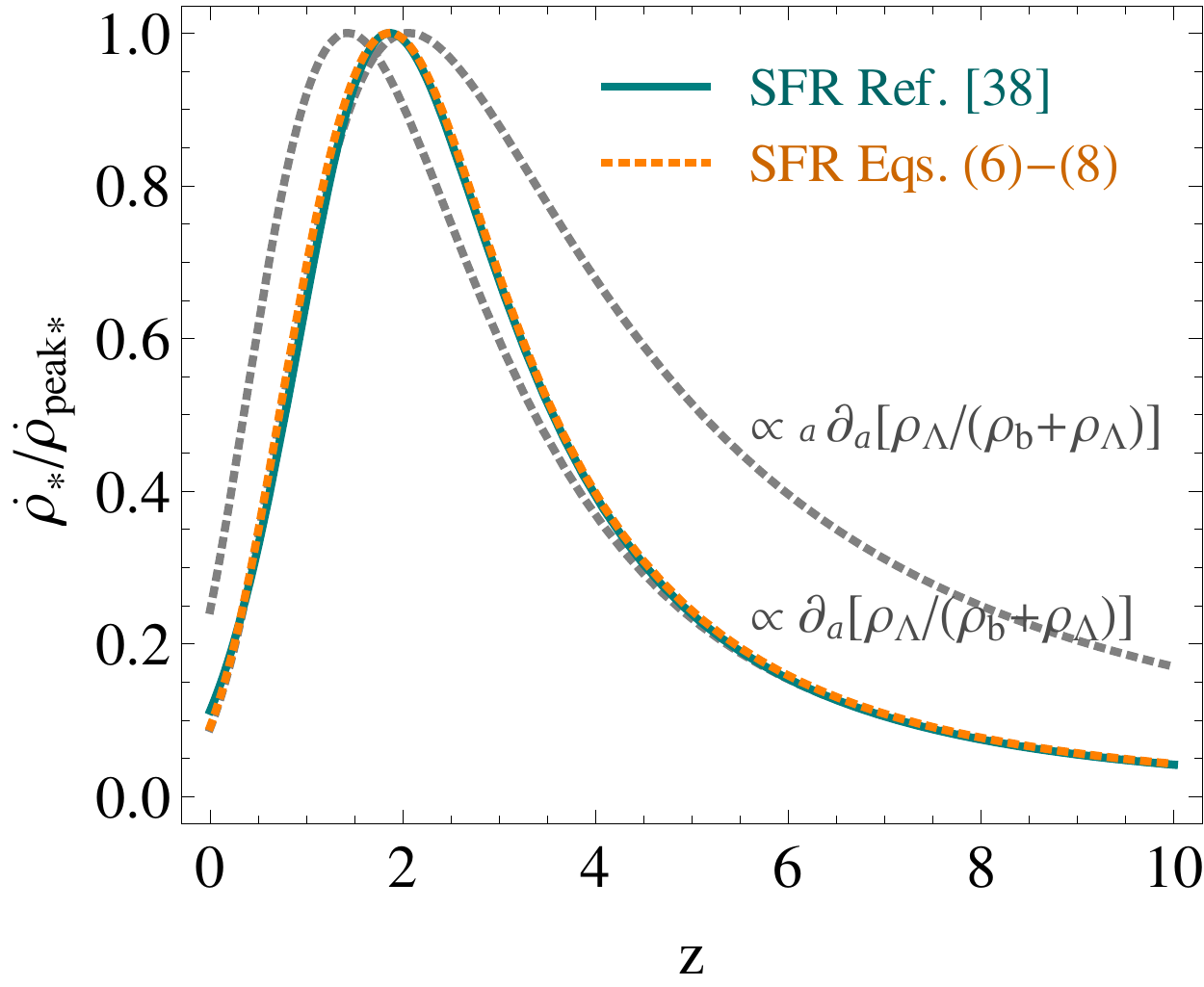}}
 \resizebox{0.3274\hsize}{!}{\includegraphics{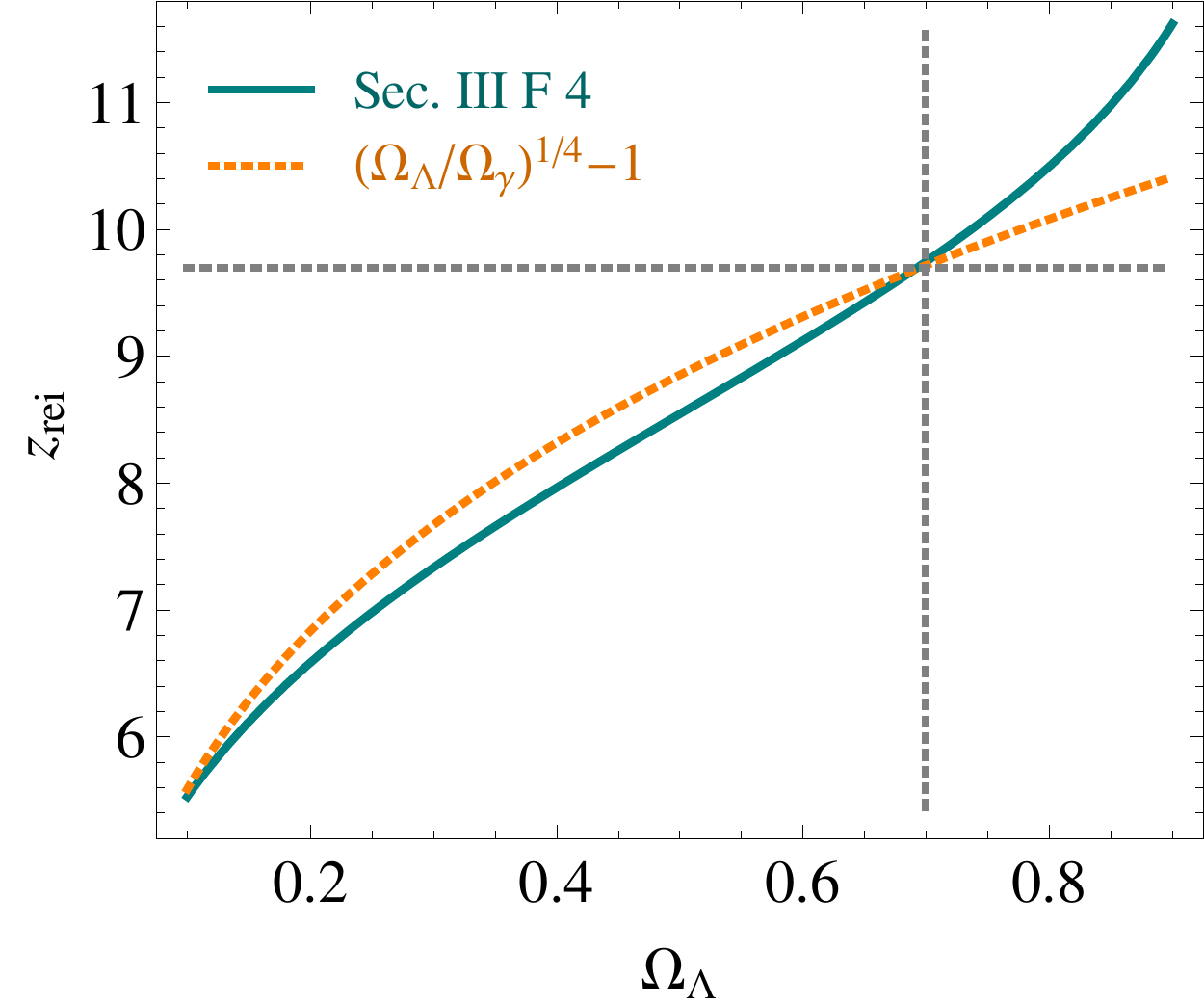}}
 \resizebox{0.3274\hsize}{!}{\includegraphics{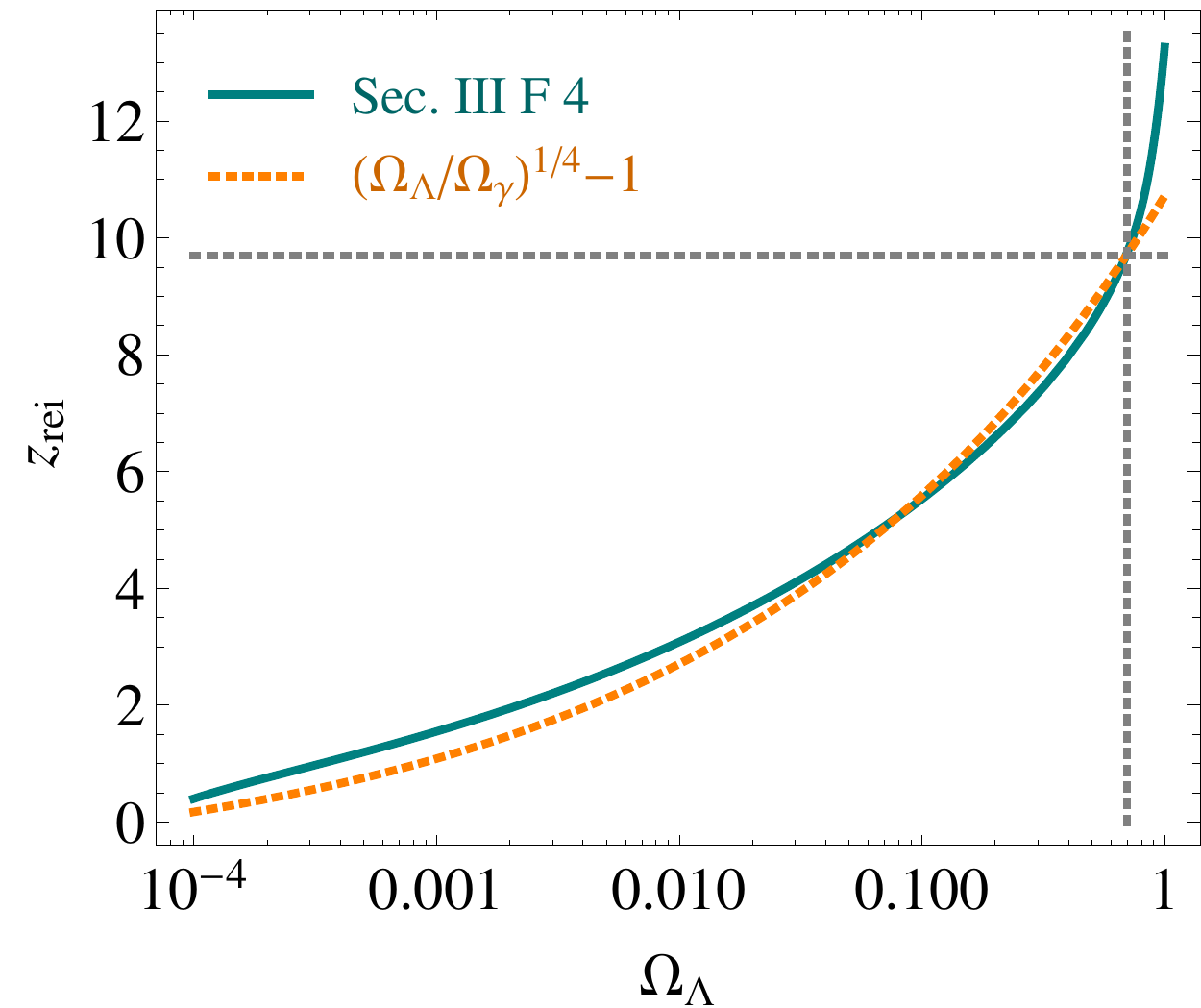}}
 \caption{\emph{Left panel:} Star-formation rate normalized at its peak from empirical results of Ref.~\cite{Madau:2014bja} and from Eqs.~\eqref{eq:f1} to \eqref{eq:sfrapp}.
 \emph{Middle and right panels:} Reionization redshift $z_{\rm rei}$ from approximated star-formation history and empirical reionization condition as a function of the cosmological constant energy density parameter $\Omega_{\Lambda}$ and from the concurrence hypothesis $z_{\rm rei}=(\Omega_{\Lambda}/\Omega_{\gamma})^{1/4}-1$.
 The overlap in the wide range of $\Omega_{\Lambda}$ values suggests that the concurrence at reionization may arise from a strong connection between the star-formation rate and the cosmological background dynamics rather than a fundamental physical connection between $\Omega_{\Lambda}$ and $\Omega_{\gamma}$.
 }
 \label{fig:zcoinc}
\end{figure*}

As a motivation for more rigorous future work, we shall give a brief and simplified exploration of how a connection between star formation and the cosmological background dynamics may give rise to scenario (i) in Sec.~\ref{sec:interpretation}.
For this purpose, we first consider the empirical star-formation rate of Ref.~\cite{Madau:2014bja},
\begin{equation}
 \frac{\dot{\rho}_*}{\rho_{\rm c0}} \propto \frac{(1+z)^{2.7}}{1+\left(\frac{1+z}{2.9}\right)^{5.6}} \,,
 \label{eq:sfr}
\end{equation}
where $\rho_{\rm c0}$ is the critical energy density today.
In Fig.~\ref{fig:coincidences}, we observed an interesting correlation between the transition in relative dominance of $\rho_{\Lambda}$ and $\rho_{\rm b}$ with the star-formation peak.
More specifically, this is close to where the transition rate is maximal, Eq.~\eqref{eq:lambdasfr}.
The transition rate, or the first derivative in Eq.~\eqref{eq:lambdasfr}, also closely matches the decline in the star-formation rate at $z\lesssim2$, whereas the high-redshift tail is well described by $(1+z)^{-1}$ times this derivative.
We show a comparison of these terms with normalization at their peaks against a normalized Eq.~(\ref{eq:sfr}) in Fig.~\ref{fig:zcoinc}, adopting \emph{Planck} cosmological parameters.

The match is remarkable and
we shall for now assume that this is indicative of a physical connection.
As a rough inductive approximation for a cosmology dependent star-formation rate serving our brief exploration, we therefore adopt the normalized inner and outer functional behaviors from the derivatives,
\begin{eqnarray}
 f_{\rm in} & = & \frac{1}{2} \left(\frac{3}{2}\right)^2 \left( \frac{\Omega_{\rm b}}{2\Omega_{\Lambda}} \right)^{1/3} (1+z) f_{\rm out} \,, \label{eq:f1} \\
 f_{\rm out} & = & \frac{4 \Omega_{\rm b} \Omega_{\Lambda} (1+z)^3 }{\left[ \Omega_{\rm b}(1+z)^3 + \Omega_{\Lambda}\right]^2} \,. \label{eq:f2}
\end{eqnarray}
We then match the two functions with a weighting function $w(z)$ to interpolate between the distinct rates at low and high redshifts.
More specifically, we use the interpolation
\begin{equation}
 \frac{\dot{\rho}_*}{\rho_{\rm c0}} \propto \frac{ f_{\rm in} + w\,f_{\rm out}}{1+w} \,.
 \label{eq:sfrapp}
\end{equation}
The weighting function introduces some degree of freedom, where however the particular choice is not crucial for the following discussion as long as the interpolation happens relatively quickly (see Fig.~\ref{fig:zcoinc}).
We use the simple function $w(z)=[(1+z)/(1+z_{\rm infl})]^n$, where $z_{\rm infl}$ denotes the inflection redshift of $f_{\rm out}$ given by $(1+z_{\rm infl})=[(3+\sqrt{7})\Omega_{\Lambda}/(2\Omega_{\rm b})]^{1/3}$. As we want a swift transition we pick $n=7$.
The resulting approximation for the star-formation rate in Eq.~\eqref{eq:sfrapp} normalized at its peak is shown in Fig.~\ref{fig:zcoinc} and is remarkably accurate.

We emphasize however that while it gives a very good match to Eq.~\eqref{eq:sfr}, its dependence on cosmological parameters is not tested and
the inductive reasoning employed is prone to epistemic uncertainty.
It is worth noting however that the qualitative effects on the star-formation rate with changes of $\Omega_{\Lambda}$ are consistent with the theoretical model of Ref.~\cite{Bousso:2008bu} in that an increase and decrease in $\Omega_{\Lambda}$ at fixed baryonic matter fraction moves the star-formation rate to earlier and later times, respectively, while the normalized shape and peak of the curve remains relatively robust at early times for increased $\Omega_{\Lambda}$.
The dependence of the star-formation rate on $\rho_{\Lambda}$ enters through the virialization of the halo that sets the timescale for interior dynamical processes such as the baryon cooling time preceding star formation, which also depends on the baryonic fraction~\cite{Bousso:2008bu}.
The connection to the background dynamics is introduced here with the halo virialization.

Next, we introduce an empirical condition for half of the reionization to have occurred by a redshift $z_{\rm rei}$.
For this, we simply evaluate $\rho_*(z_{\rm rei})$ and relate it to the constraint $\rho_{*0}=10^{-1}\rho_{{\rm b}0}$~\cite{Nagamine:2006kz,Bousso:2008bu} with $\rho_{*0}$ denoting the integrated energy density that went into stars.
This yields approximately $\rho_*(z_{\rm rei})\approx 10^{-2} \rho_{*0}$ for the \emph{Planck} cosmology. 
We use the standard configuration concurrence model results (Table~\ref{tab:constraints}) to fix $\rho_*(z_{\rm rei})/\rho_{*0}=1.24\times10^{-2}$.
We then fix all the cosmological parameters besides $\Omega_{\Lambda}$ and use Eq.~\eqref{eq:sfrapp} to find $z_{\rm rei}$ as a function of $\Omega_{\Lambda}$ using our empirical reionization condition.
The resulting reionization redshifts are shown in Fig.~\ref{fig:zcoinc}.
Note that the limit of $\Omega_{\Lambda}\rightarrow1$ implies that $\Omega_{\rm c}\rightarrow-(\Omega_{\rm b}+\Omega_{\rm rad})$ with massless neutrinos.

Interestingly, we find that the relation adopted for the concurrence model, Eq.~\eqref{eq:zcoinc}, i.e., $(1+z_{\rm rei})=(\Omega_{\Lambda}/\Omega_{\gamma})^{1/4}$, provides a very good match to this result for $\Omega_{\Lambda}\lesssim0.8$.
Note that $\Omega_{\gamma}$ does not appear in Eq.~\eqref{eq:sfrapp}. This may imply that it only coincidentally provides the correct normalization for $z_{\rm rei}(\Omega_{\Lambda})\propto\Omega_{\Lambda}^{1/4}$, although possibly slightly high in comparison to astrophysical constraints (Sec.~\ref{sec:intro}).
Importantly, the peak of matter domination is given by a similar expression involving $\Omega_{\gamma}$ (Sec.~\ref{sec:concerns}).
Within scenario (i) of Sec.~\ref{sec:interpretation}, the normalization with $\Omega_{\gamma}$ would suggest a connection to the comparable values of $\rho_{\rm b}$ and $\rho_{\gamma}$ at recombination with $\Omega_{\gamma}=\Omega_{\rm b}(1+z_{\rm rec})^{-1}\sim10^{-3}\Omega_{\rm b}$ such that $(\Omega_{\Lambda}/\Omega_{\gamma})^{1/4}\sim5(\Omega_{\Lambda}/\Omega_{\rm b})^{1/4}$, entering through the reionization condition.

While only approximative and inductive, our exploration suggests that the reionization coincidence may have an origin in the similarity of the star-formation rate to cosmological background dynamics.
The connection of the cosmological constant with the star-formation history therefore seems worthwhile pursuing (e.g.,~\cite{Bousso:2008bu,Loeb:2016vps}).
We leave a more thorough analysis supported by simulations and physical star-formation models to future work.

\section{Conclusions} \label{sec:conclusions}

Unraveling the nature underlying the observed late-time accelerated expansion of our Universe is one of the prime endeavors in cosmology.
The observation that the energy densities of the cosmological constant, or dark energy, and matter are of comparable size today may plausibly be an important guide  to finding a solution to this puzzle.
Another well-known cosmic coincidence is the concurrence of the matter-radiation equality with recombination.
In this paper, we point out for the first time that moreover the third equality, between the energy densities of radiation and the cosmological constant, coincides with the epoch of reionization.
To quantify the statistical relevance of this new coincidence, we compute the Bayes factor between the concordance cosmology and a model which imposes a match between the time of equality and the reionization epoch.
We find a \emph{very strong} preference for such a concurrence on the Jeffreys scale.
We furthermore find limited impact from the choice of prior on the optical depth, changes in the duration of reionization, or from allowing for a free sum of neutrino masses. 
The preference of the concurrence remains \emph{substantial} with the inclusion of approximate constraints from intermediate \emph{Planck} HFI data.
For two forecasts of SKA 21~cm data centering around the current mean values of the optical depth, we find the potential for \emph{substantial} or \emph{decisive} preference against or in favor of the concurrence model, respectively.

We compare the new coincidence with the traditional coincidence problems.
In particular, we find a closer relative proximity in redshift and age compared to the other cosmological coincidences and a narrower range for possible concurrences of the energy density equalities with cosmic epochs or events from the interchange of matter or radiation components. 
We discuss some concerns and caveats around the analysis but also provide possible physical interpretations of a correlation.
In particular, we explore how a connection between the star-formation rate and the cosmological background dynamics could give rise to the concurrence between the radiation to cosmological constant equality and the epoch of reionization.

With our results and discussion provided we ultimately leave it up to the reader to decide whether the new coincidence at the reionization epoch should indeed be considered a problem or not.
From our brief investigation, we however conclude that the reionization coincidence with the cosmological constant and radiation energy density equality is at least not lesser of a problem than the traditional cosmological coincidence problems.


\begin{acknowledgments}
We thank Luke Butcher, Alan Heavens, John Peacock, Yabebal Tadesse, Eric Tittley, and Anna Lisa Varri for useful discussions and help with \textsc{MCEvidence}.
L.L.~was supported by a SNSF Advanced Postdoc.Mobility Fellowship (No.~161058) and the STFC Consolidated Grant for Astronomy and Astrophysics at the University of Edinburgh.
V.S.-B.~acknowledges funding provided by CONACyT and the University of Edinburgh.
Numerical computations were conducted on the COSMOS Shared Memory system at DAMTP, University of Cambridge operated on behalf of the STFC DiRAC HPC Facility. This equipment is funded by the BIS National E-infrastructure capital grant ST/J005673/1 and STFC grants ST/H008586/1, ST/K00333X/1.
Potential subliminal inspiration drawn from the late works of Charles Piazzi Smyth cannot be excluded.
Please contact the authors for access to research materials.
\end{acknowledgments}


\appendix
\section{\emph{Planck} reionization history and modifications to \textsc{CosmoMC}} \label{sec:planckreion}

For completeness, we provide here a brief review of the reionization history used in the \emph{Planck} analysis~\cite{Ade:2015xua} and adopted in this work.
We follow the discussion presented in Refs.~\cite{Lewis:2008wr,camb_notes}.

The optical depth is given by
\begin{equation}
 \tau \equiv \int_{\eta_*}^{\eta_0} d\eta \, a \, n_e^{\rm rei} \sigma_T \,,
 \label{eq:tau}
\end{equation}
where $\eta_0$ is the present conformal time, $\eta_*$ the conformal time at last scattering, and $\sigma_T$ is the Thomson scattering cross section.
Furthermore, $n_e^{\rm rei}$ denotes the number density of electrons freed by reionization.
It dominates the current electron number density $n_e$ such that $n_e\approx n_e^{\rm rei}$.
Eq.~\eqref{eq:tau} can be rewritten as $\tau \propto \int dy \, x_e(y)$, where $x_e\propto n_e(1+z)^{-3}$ denotes the number of free electrons per hydrogen atom, $y\equiv(1+z)^{3/2}$, and we have assumed reionization during matter domination.

The reionization history is modeled by the function
\begin{equation}
 x_e(y) = \frac{f}{2} \left[ 1 + \tanh\left( \frac{y(z_{\rm rei})-y}{\Delta_y} \right) \right] \,,
\end{equation}
where $f$ is the maximal reionization fraction and $\Delta_y$ is a width that is related to a redshift width in the reionization history $\Delta_y=1.5\sqrt{1+z_{\rm rei}}\Delta_z$ with default value $\Delta_z=0.5$ (see Sec.~\ref{sec:robustness} for a change to $\Delta_z=1.5$).
As pointed out in Sec.~\ref{sec:parameters}, where $\Delta_y$ remains matter-dominated, $z_{\rm rei}$ is approximately the redshift of instantaneous reionization for the same optical depth.
Importantly, it is the optical depth $\tau$ that affects the CMB (see Sec.~\ref{sec:datasets}).
In Sec.~\ref{sec:robustness}, we have tested the robustness of this approximation by widening $\Delta_z$.
Finally, helium ionization also needs to be accounted for, which changes the reionization fraction to $f\sim1.08$, obtained from the helium mass fraction $Y_{\rm {P}}$, and furthermore adds a transition in the reionization history at $z\sim3.5$, which affects $\tau$ at the level of $\sim0.001$~\cite{Lewis:2008wr}.

In order to implement the coincidence model in \textsc{CosmoMC}, we fix $\tau$ and $z_{\mathrm{rei}}$ at arbitrary values in the sampler and overrun the code with an iteration in $\tau$ around Eq.~(\ref{eq:zcoinc}). 
\textsc{CAMB} employs a binary search to map the optical depth to the reionization redshift, which we adopt in this process until we reach sub-percent level agreement in the coincidental redshifts, i.e., well within the $1\sigma$ region of $z_{\rm rei}$ (see Table~\ref{tab:constraints}).
The resulting optical depth and reionization values are then stored as derived parameters.

Finally, we note that a simple approximation for the optical depth, assuming instantaneous reionization, is given by~\cite{Shull:2008su,Liu:2015txa,Aghanim:2016yuo}
\begin{equation}
 \tau \approx \frac{2 \sigma_{\rm T}(1-Y_{\rm P})\,H_0}{8\pi G \,m_p} \frac{\Omega_{\rm b}}{\Omega_{\rm m}} \left[ \sqrt{\Omega_{\rm m}(1+z_{\rm rei})^3 + \Omega_{\Lambda}} - 1 \right] \,,
 \label{eq:tauapprox}
\end{equation}
which holds within the $\sigma_{\tau}$ of Table~\ref{tab:constraints}.
Eq.~\eqref{eq:tauapprox} can be used with Eq.~\eqref{eq:zcoinc} to directly infer $\tau$ from the cosmological parameters of a concurrence model, which roughly yields
$\tau\propto h\Omega_{\rm b}\Omega_{\rm m}^{-1/2} (\Omega_{\Lambda}/\Omega_{\gamma})^{3/8}$.


\section{Further views of the coincidences} \label{sec:furtherviewsofcoincidences}

The \emph{Why Now?} problem of the cosmological constant 
can be more strikingly formulated if considering the inflection point of the relative dominance between the energy densities of total matter and the cosmological constant.
For \emph{Planck} cosmology (Sec.~\ref{sec:coincidence}), this yields
\begin{equation}
 \partial_z^2\left( \frac{\rho_{\Lambda}}{\rho_{\rm m}+\rho_{\Lambda}} \right) = 0 \ \ \textrm{at} \ \  z = -1 + \left( \frac{\Omega_{\Lambda}}{2\Omega_{\rm m}} \right)^{1/3} \approx 0.0 \,. \label{eq:ccderivativecoinc}
\end{equation}
Comparably, one finds
\begin{align}
 \partial_a^2\left( \frac{\rho_{\Lambda}}{\rho_{\rm b}+\rho_{\Lambda}} \right) = 0 & \ \ \textrm{at} \ \ z = -1 + \left( \frac{2\Omega_{\Lambda}}{\Omega_{\rm b}} \right)^{1/3} \approx 2.1 \,, \ \ \label{eq:lambdasfr} \\
 \partial_z^2\left( \frac{\rho_{\Lambda}}{\rho_{\gamma}+\rho_{\Lambda}} \right) = 0 &  \ \ \textrm{at} \ \  z = -1 + \left( \frac{3\Omega_{\Lambda}}{5\Omega_{\gamma}} \right)^{1/4} \approx 8.4 \,.
\end{align}
Since including derivatives implies the introduction of extra degrees of freedom that can produce coincidental redshifts, we have only considered the intersection points of $\rho_i$ in our analysis.
Note however that we have used the first derivative in Eq.~\eqref{eq:lambdasfr} to describe an approximate star-formation rate in Sec.~\ref{sec:sfr}.


\vfill
\bibliographystyle{arxiv_physrev.bst}
\bibliography{reioncoinc.bib}

\end{document}